\documentstyle[sprocl]{article}

\title{
\begin{flushright}
{\normalsize Yaroslavl State University\\
             Preprint YARU-HE-97/07\\
             hep-ph/9710219} \\[10mm]
\end{flushright}
Neutrino transitions $\nu \to \nu \gamma$,
       $\nu \to \nu e^+ e^-$ in a strong magnetic field 
       as a possible origin of cosmological $\gamma$-burst} 
 
\author{A.A.~Gvozdev$^{1}$, A.V.~Kuznetsov$^{1}$, 
        N.V.~Mikheev$^{1}$, L.A.~Vassilevskaya$^{1,2}$}

\address{
     1. Yaroslavl P.G.~Demidov State University, Yaroslavl, Russia \\
     2. Moscow M.V.~Lomonosov State University, Moscow, Russia}

\date{}

\begin{document}

\pagestyle{empty} 

\maketitle

\begin{abstract}
The high energy neutrino transitions with 
  the photon and electron-positron pair creation  
 in a strong magnetic field  in the framework 
of the Standard Model are investigated. The process probabilities 
 and the mean values 
of the neutrino energy and momentum loss are presented. 
The asymmetry of outgoing neutrinos, as a possible source of sufficient 
recoil ``kick'' velocity of a remnant and the emission of $e^+ e^-$-pairs
and $\gamma$-quanta in a ``polar cap'' region of a remnant, as a possible 
origin of cosmological $\gamma$-burst are discussed.  
\end{abstract}

\vspace{20mm}

\begin{center}
{\it Talk presented at the International Workshop \\
on Non-Accelerator New Physics (NANP-97), \\
Dubna, Russia, July 7-11, 1997}
\end{center}

\newpage

\section{Introduction}

The absorption, emission, and scattering of neutrinos are of great 
importance in astrophysics. These combined processes occurs in a 
star medium. Hot and dense star matter or plasma is usually 
considered as the medium. Of particular conceptual interest are those 
neutrino reactions in dense matter (or plasma) which are forbidden 
or suppressed substantially in a vacuum. Thus under favorable conditions 
the conversion of one neutrino flavor into another is greatly enhanced by 
the presence of the background~\cite{MS}. 
 Some time ago it was observed~\cite{OS} that the 
electromagnetic properties of neutrinos are also drastically modified 
within a dense matter when compared to the properties in vacuum. At the 
present time the neutrino propagation in magnetized dense matter is the 
subject of intensive investigations~\cite{SV}. The effect of the background 
matter upon the neutrino transitions can lead to such interesting phenomena 
as the Cherenkov radiation by massless neutrinos~\cite{NP} and the possible 
explanation of pulsar ``kick'' velocities by neutrino oscillations biased by  
a magnetic field~\cite{KG}. 

The Cherenkov radiation $\nu(p) \to \nu(p') + \gamma(q)^*$  by massless 
neutrinos as well as the plasmon decay into $\nu \bar\nu$-pair~\cite{RAF} is
 forbidden in vacuum. In the presence of a matter however neutrinos 
acquire an effective $\nu\nu\gamma$-coupling and the processes are 
kinematically allowed because the photon acquire essentially an effective 
mass square (positive or negative). 

We stress that a strong magnetic field itself can also play the role 
of an active medium and in the presence of the magnetic field neutrinos 
also acquire an effective $\nu\nu\gamma$-coupling just as in a dense 
matter. 
It was observed that  a field-induced amplitude of the massive 
neutrino radiative decay $\nu_i \rightarrow \nu_j + \gamma$ 
\footnote{Here $i$ and $j$ enumerate the neutrino mass eigenstates but not 
the neutrinos with definite flavors.} was not suppressed by the smallness 
of neutrino masses and did not vanish even in the case of the massless 
neutrino as opposed to the vacuum amplitude~\cite{GMV,GMV96}. 
The photon dispersion in the strong magnetic field 
differs significantly from the vacuum dispersion 
with increasing of the photon energy~\cite{Ad}, so 
the on-shell photon  4-momentum can appear as the space-like  
with sufficiently large value of $q^2$, 
($\vert q^2 \vert \gg m^2_\nu$). 
In this case the phase space for the neutrino transition 
$\nu_i \rightarrow \nu_j + \gamma$ with $m_i < m_j$ is opened also. 
It means that the decay probability of ultrarelativistic neutrino 
becomes insensitive 
to the neutrino mass spectrum due to the photon dispersion relation 
in the strong magnetic field. This phenomenon results in a strong 
suppression ($ \sim m^2_\nu / E^2_\nu$) of the neutrino transition 
with flavour violation, so a diagonal process 
$\nu_l \rightarrow \nu_l + \gamma$ ($l = e, \mu, \tau$) 
is realized only. Thus, this diagonal radiative neutrino transition 
does not contain 
uncertainties associated with a possible mixing in the lepton sector 
of the Standard Model, and can lead to observable physical effects in  
the strong magnetic fields. 

Another decay channel also exists, $\nu_i \to \nu_j e^- e^+$, which is
forbidden in vacuum when $m_i < m_j + 2m_e$. 
However, the kinematics of a charged particle 
in a magnetic field is that which allows to have 
a sufficiently large space-like total momentum for 
the electron-positron pair, and this process is possible even for very light 
neutrinos. It means that a flavor of the ultrarelativistic neutrino is also 
conserved in this transition in a magnetic field, to the terms of the order 
of $m^2_\nu/E^2_\nu$ regardless of the lepton mixing angles. 
Consequently, a question of neutrino 
mixing is not pertinent in this case and the process 
$\nu \to \nu e^- e^+$ can be considered in the frame of 
the Standard Model without lepton mixing.

Here we investigate the processes $\nu \to \nu \gamma$ and 
$\nu \to \nu e^- e^+$ for the high 
energy neutrino, $E_\nu \gg m_e$, in a strong constant magnetic field. 
We consider a magnetic field as the strong one 
if it is much greater than the known Schwinger value 
$B_e = m^2_e/e \simeq 4.41 \cdot 10^{13} G$. 
These processes could be of importance in astrophysical applications, 
e.g. in an 
analysis of cataclysms like a supernova explosion or a coalescence of 
neutron stars, where the strong magnetic fields can exist, and where 
neutrino processes play the central physical role. At the present
time there is some evidance that neutron stars with unusually strong 
magnetic field strength $B \sim 10^{15} \div 10^{16}G$ and even up to 
$3 \cdot 10^{17}G$ both for toroidal~\cite{Lipu} and for 
poloidal~\cite{Bocq} 
fields can exist in nature. Such high-field neutron stars 
(so-called ``magnetars'') known as a model of SGRR -- pulsars~\cite{DT}.     

Here we calculate the probabilities of the processes and the mean 
values of the neutrino energy and momentum loss through the production 
of electron-positron pairs and photons in such a ``magnetar''. 

If the momentum transferred is relatively small, $|q^2| \ll m^2_W$
\footnote{As the analysis shows, it corresponds in this case to the 
neutrino energy $E \ll m^3_W/e B$.}, 
the weak interaction of neutrinos with electrons could be described 
in the local limit by the effective Lagrangian of the form

\begin{eqnarray}
{\cal L} \, = \, \frac{G_F}{\sqrt 2} 
\big [ \bar e \gamma_\alpha (g_V - g_A \gamma_5) e \big ] \,
\big [ \bar \nu \gamma^\alpha (1 - \gamma_5) \nu \big ] , 
\label{eq:L} \\
g_V = \pm {1 \over 2} + 2 \sin^2 \theta_W , \quad g_A = \pm {1 \over 2}. 
\nonumber
\end{eqnarray}

\noindent 
Here the upper signs correspond to the electron neutrino ($\nu = \nu_e$) 
when both neutral and charged current interaction takes part 
in a process. The lower signs correspond to $\mu$ and $\tau$ neutrinos 
($\nu = \nu_\mu, \nu_\tau$), when the neutral current interaction 
is only presented in the Lagrangian~(\ref{eq:L}). 

The strong magnetic field is the only exotic we use. 

\section{The probability of the process $\nu \to \nu \gamma$}

The field-induced $\nu\nu\gamma$-vertex
can be calculated using an effective four-fermion
weak interaction~(\ref{eq:L}) 
of the left neutrino with the electron only, because
the electron is the most sensitive fermion to the external field.
By this means, the diagrams describing this process are reduced to an 
effective diagram with the electron in the loop.

The calculation technique for the loop diagram of this type was 
described in detail in the paper~\cite{GMV96}.
We note that this effective $\nu\nu\gamma$ amplitude 
is enhanced substantially in the vicinity of the so-called  
photon cyclotronic frequencies. The same phenomenon in the field-induced 
vacuum polarization is known as the cyclotronic resonance~\cite{Sh2}.

As was first shown by Adler~\cite{Ad}, two eigenmodes of the photon 
propagation with polarization vectors

\vspace{-5mm}

\begin{eqnarray}
\varepsilon _{\mu}^{(\parallel)} & = & 
\frac{ (q \varphi)_{\mu} }{ \sqrt{ 
 q^2_\perp  } } ; \; \; \; \; \;
\varepsilon _{\mu}^{(\perp)} = \frac{ (q \tilde
\varphi)_{\mu} }{ \sqrt{ q^2_\parallel } }
\label{eq:EP}
\end{eqnarray}

\noindent 
are realized in the magnetic field, the so-called parallel ($\parallel$) 
and perpendicular ($\perp$) polarizations (Adler's notations). 
Here $\varphi_{\alpha \beta} = F_{\alpha \beta} / B$ and 
${\tilde \varphi}_{\alpha \beta} = \frac{1}{2} \varepsilon_{\alpha \beta
\mu \nu} \varphi_{\mu \nu} \; $ are the dimensionless tensor and dual
tensor of the external magnetic field with the strength $\vec B = (0,0,B)$;
$q^2_{\parallel}  =  ( q \tilde \varphi \tilde \varphi q ) =
q_\alpha \tilde \varphi_{\alpha\beta} \tilde \varphi_{\beta\mu} q_\mu 
= q^2_0 - q^2_3$,
$q^2_{\perp}  =  ( q \varphi \varphi q ) = q^2_1 + q^2_2$. 

It is of interest for some astrophysical applications
the case of relatively high neutrino energy 
$ E_\nu \simeq 10 \div 20 MeV \gg m_e$ 
and strong magnetic field $e B > E^2_\nu$. 
As the analysis of the photon dispersion in a strong magnetic field shows, 
a region of the cyclotronic resonance $q^2_\parallel \sim 4 m_e^2$ 
in the phase space of the final photon, 
corresponding to the ground Landau level of virtual electrons dominates 
in the process $\nu \to \nu \gamma$. 
It is particularly remarkable that the $\perp$ photon mode only acquires 
a large space-like 4-momentum in the vicinity of the resonance. The 
corresponding amplitude of the process contains the enhancement due to 
the square-root singularity when $q^2_\parallel \to 4 m_e^2$. 
It means that taking account of the many-loop radiative corrections is 
needed. A detailed description of this procedure will be published 
elsewhere~\cite{VGM}. 
A general expression of the probability of the process 
$\nu \to \nu \gamma^{(\perp)}$ has a rather complicated form. 
Here we present the result of our calculation in the limit 
$e B \gg E^2_\nu \sin^2 \theta$: 

\vspace{-3mm}

\begin{equation} 
W^{(\gamma)} \simeq \frac{\alpha G^2_F}{8 \pi^2 }(g^2_V + g^2_A) 
e^2 B^2 E \sin^2{\theta}.
\label{eq:w1}
\end{equation}

\noindent 
Here $E$ is the initial neutrino energy, 
$\theta$ is the angle between the vectors 
of the magnetic field strength $\vec B$ and the momentum of
the initial neutrino $\vec p$. The dispersion relation for the 
$\parallel$ photon mode is close to the vacuum one, $q^2 = 0$, and it gives 
a negligibly small contribution to the probability in the limit considered.

The probability of the process $\nu \to \nu \gamma^{(\perp)}$ is also 
non-zero above the threshold point of the $e^- e^+$-pair creation, 
$q^2_\parallel > 4 m_e^2$, due to an imaginary part of the amplitude. 
However, the $\perp$-photon mode is unstable and another channel 
$\nu \to \nu e^- e^+$ dominates in this region. 

\section{The probability of the process $\nu \to \nu e^- e^+$}

An amplitude of the process $\nu \to \nu e^- e^+$ could be immediately 
obtained from the Lagrangian~(\ref{eq:L}) where the known solutions of 
the Dirac equation in a magnetic field should be used. 
We present here the results of our calculations of the probability 
in the strong field limit $e B \gg E^2 \sin^2 \theta$ which is of
more importance in some astrophysical applications.

In the case when the field strength $B$ appears to be 
the largest physical parameter, the electron and the positron could 
be born only in the states corresponding to the lowest Landau level. 
Integrating over the phase space, one obtains the following 
expression for the probability in the limit $e B \gg E^2 \sin^2 \theta$

\begin{equation}
W^{(e e)} = \frac{G_F^2 (g_V^2 + g_A^2)}{16 \pi^3} \, 
e B E^3 \sin^4 \theta .
\label{eq:w2}
\end{equation}

\section{The neutrino energy and momentum losses}

It should be noted that a practical significance of these processes for 
astrophysics could be in the mean values of the neutrino energy and 
momentum losses rather than in the process probabilities. 
These mean values could be found from the four-vector 

\begin{equation}
Q^\alpha \, = \, E \int d W q^\alpha \, = \, ({\cal I}, \vec {\cal F}) E.
\label{eq:Q}
\end{equation}

\noindent 
Its zero component is connected with the mean neutrino energy 
loss in a unit time, ${\cal I} = dE/dt$.
The space components of the four-vector~(\ref{eq:Q}) are connected 
similarly with the neutrino momentum 
loss in unit time, $\vec {\cal F} = d\vec p/dt$.
Here we present the expressions for $Q^{\alpha}$ in the strong
field limit for both processes:

\vspace{-3mm}

\begin{eqnarray}
{\cal I} & = & E W \, C_1 \, 
\left(1 + \frac{2 g_V g_A}{g_V^2 + g_A^2} \cos\theta\right),
\label{eq:I} \\[2mm]
{\cal F}_z & =  & E W \, C_1 \, 
\left(\cos\theta  + \frac{2 g_V g_A}{g_V^2 + g_A^2}\right), \quad
{\cal F}_\perp \, =  \, E W \, C_2 \, \sin\theta,
\label{eq:F} 
\end{eqnarray}

\noindent 
where the $z$ axis is directed along the field, the vector 
$\vec {\cal F}_\perp$, transverse to the field, lies in the plane of 
the vectors $\vec B$ and $\vec p$,

\vspace{-3mm}

\begin{displaymath}
C_1^{(\gamma)} = \frac{1}{4}, 
\quad 
C_2^{(\gamma)} = \frac{1}{2}, 
\quad 
C_1^{(ee)} = \frac{1}{3}, 
\quad 
C_2^{(ee)} = 1. 
\end{displaymath}

We emphasize  that all expressions for the processes 
($\nu \to \nu e^- e^+$, $\nu \to \nu \gamma$)
are applicable for the processes with antineutrino due to the 
$CP$-invariance of the weak interaction.

It should be mentioned also that our results are valid in the presence 
of plasma with the electron density $n \sim 10^{33} \div 10^{34} cm^{-3}$  
($n = n_{e^-} - n_{e^+}$). This is due to a peculiarity of the statistics 
of the relativistic electron gas in a magnetic field~\cite{Lan}.  
As the analysis shows, 
the suppressing statistical factors in integrating over the phase 
space of the final particles do not arise at the conditions
$B > 10^{15} \; G \; (T/3~{MeV})^2$ and
%
%\noindent 
$B > 5 \cdot 10^{15} \; G \; (n/10^{33}~cm^{-3})^{2/3}$. 

\section{Possible astrophysical consequences}

To illustrate a possible application of our results we estimate 
below the neutrino energy and momentum losses in some astrophysical  
cataclysm of type of a supernova explosion or a merger of 
neutron stars. We assume that for some reasons a compact remnant 
has a very strong poloidal magnetic field, 
$B \sim 10^{15} \div 10^{17} \; G$. Objects of such a type, 
so-called ``magnetars'', were investigated in the paper~\cite{DT}. 
According to standard astrophysical models~\cite{Imsh} 
the neutrinos of all species with the typical mean energy 
$\bar E_\nu \sim 20 MeV$ are radiated from a neutrinosphere 
in above mentioned astrophysical cataclysm. 
The electron density in the vicinity of neutrinosphere will be  
considered to be not too high, so a creation of the $e^- e^+$ 
pairs is not suppressed by statistical factors. 
In this case the neutrino propagating through the 
magnetic field will loose the energy and the momentum in accordance 
with our formulas. A dominant contribution to the total energy lost by 
neutrinos in the field due to the process of the $e^- e^+$ pair 
creation could be estimated from Eq.~(\ref{eq:I}):

\vspace{-3mm}

\begin{equation}
\frac{\Delta {\cal E}^{(ee)}}{{\cal E}_{tot}} 
  \sim  10^{-2}\, 
\left (\frac{B}{10^{17} \, G} \right )\,
\left (\frac{\bar E}{20 \, MeV} \right )^3\,
\left (\frac{ \Delta \ell}{10 \, km} \right ), 
\label{eq:DE1}
\end{equation}

\noindent 
here $\Delta \ell$ is a characteristic size of the region  
where the field strength varies in\-signi\-fi\-cant\-ly, 
${\cal E}_{tot}$ is the total energy carried off by 
neutrinos in a supernova explosion, 
$\bar E$ is the neutrino energy averaged over the neutrino spectrum. 
Here we take the energy scales which are believed to be typical for 
supernova explosions~\cite{Imsh}. 

An asymmetry of outgoing neutrinos is another interesting manifestation 

\begin{equation}
A \; = \; \frac{|\sum_i {\bf p}_i|}{\sum_i |{\bf p}_i|}.
\label{eq:A1}
\end{equation}

\noindent 
In the same limit of the strong field we obtain 

\begin{equation}
A^{(ee)}  \sim  10^{-2}\, 
\left (\frac{B}{10^{17} \, G} \right )
\left (\frac{\bar E}{20 \, MeV} \right )^3 
\left (\frac{ \Delta \ell}{10 \, km} \right ).
\label{eq:A2}
\end{equation}

\noindent  
One can see from Eqs.~(\ref{eq:DE1}), (\ref{eq:A2}) that the effect could 
manifest itself at a level of about percent. In principle, it could be 
essential in a detailed theoretical description of the process of 
supernova explosion. 
For the process $\nu \to \nu \gamma$ one obtains 

\begin{equation}
A^{(\gamma)}  \sim  2 \pi \alpha \frac{e B}{E^2} A^{(ee)}. 
\label{eq:A3}
\end{equation}

\noindent  
It is seen that two processes considered could be comparable for some
values of physical parameters. 

We note that $e^+ e^-$-pairs and $\gamma$-quanta produced 
are captured by a strong magnetic field and propagate along 
the field. At the first 
glance it seems that these particles are confined. However the 
magnetosphere of a ``magnetar'' has a ``polar cap'' region which is 
defined as a narrow cone along the magnetic field axis with open lines 
of the magnetic field strength. 
So the particles which are created in the mentioned above neutrino 
reactions within a narrow cone can escape outside. 
We estimate below the neutrino energy loss in a ``polar cap'' regions 
of a millisecond ``magnetar'' taking into account that the both 
processes are comparable for values of the physical parameters used: 

\begin{equation}
{\cal E} \sim 10^{48}\;  
\left (\frac{{\cal E}_{tot}}{3 \cdot 10^{53} erg} \right )
\left (\frac{B}{10^{17} G} \right )^2
% \times 
\left (\frac{\bar E}{20 MeV}\right )
\left (\frac{R}{10 km}\right )^{3} \;
\left (\frac{10^{-3} sec}{P}\right )^{2} \; erg,
%\nonumber \\
\label{eq:E} 
\end{equation}

\noindent where ${\cal E}_{tot} \sim 10^{53} \; erg$
is the typical total neutrino radiation energy; 
$P$ is the ``magnetar'' rotation period~\cite{DT}; 
$B$ is the magnetic field strength in the vicinity of 
the neutrinosphere of radius $R$. 

We pointed out that the energy loss~(\ref{eq:E}) in terms 
of $4 \pi$-geometry is close to 
deposition of enegry observed as $\gamma$-ray bursts (GRB's).
In a standard ``fireball'' model energy of order of $10^{51} \; erg$ 
is deposited in a small volume and results in an
ultrarelativistic ejecta. A collision of the ejecta with 
intergalactic medium can be a source of the GRB~\cite{DS}. 
It is interesting that a rapid rotation of the remnant combined 
with the strong magnetic field becomes popular for understanding 
GRB's afterglow~\cite{P}.

\section{Conclusions}

\noindent 
$\bullet$  
If the physical parameters would have the above-mentioned values, 
the effect of neutrino energy and momentum losses could  
manifest itself at a percent level. It could be 
essential in a detailed analysis of the process of a 
supernova explosion or merger of neutron stars. 

\noindent 
$\bullet$  
An origin of the asymmetry of the neutrino momentum loss 
with respect to the magnetic field direction is a manifestation 
of the parity violation in weak interaction (proportional to $g_V g_A$). 

\noindent 
$\bullet$  
This asymmetry results in the recoil ``kick'' velocity of a rest of the 
cataclysm. For the parameters used, it would provide a ``kick'' velocity 
of order $1000 km/s$ for a pulsar with mass of order of the solar mass. 

\noindent 
$\bullet$  
The $\gamma$ and $e^+ e^-$ emission from the ``polar cap'' region
of newborn ``magnetar'' could be observed as an anisotropic $\gamma$-burst 
with the duration of order of the neutrino emission time and of the 
energy $\sim 10^{50} \; erg$ in terms of $4 \pi$-geometry.

\section*{Acknowledgements}  
The authors are grateful to Arnon Dar for useful critical remarks. 
This work is supported in part by the International Soros Science 
Education Program under the Grants N~d97-872 (A.K.) and N~d97-900 (A.G.).

\end{document}